\documentclass[review]{elsarticle}
\usepackage{blindtext}
\usepackage{graphicx}
\usepackage[skip=\baselineskip]{caption}
\usepackage{amsmath}
\usepackage{breqn}
\usepackage{subcaption}
\usepackage{algorithm}
\usepackage[noend]{algpseudocode}
\usepackage{listings}
\usepackage{url}
\usepackage{booktabs}
\usepackage{rotating}
\usepackage{hyperref}
\usepackage{color}
\usepackage{lineno,hyperref}
\usepackage{array}
\usepackage{multirow}
\modulolinenumbers[5]
\usepackage{amssymb}
\usepackage{array, makecell}
\usepackage{boldline}
\usepackage[x11names, table]{xcolor}
\usepackage{colortbl}
\usepackage{balance}
\usepackage{mathtools}
\usepackage{color, colortbl}
\usepackage{amssymb}
\usepackage{pifont}
\newcommand{\cmark}{\ding{51}}%
\newcommand{\xmark}{\ding{55}}%
\definecolor{dkgreen}{rgb}{0,0.6,0}
\definecolor{gray}{rgb}{0.5,0.5,0.5}
\definecolor{mauve}{rgb}{0.58,0,0.82}
\definecolor{red}{rgb}{1,0,0}
\definecolor{Green}{rgb}{0,1,0}
\definecolor{yellow}{rgb}{1,1,0}
\definecolor{LightCyan}{rgb}{0.88,1,1}
\newcolumntype{L}{>{\centering\arraybackslash}m{0.8cm}} 
\newcolumntype{M}{>{\centering\arraybackslash}m{0.6cm}}
\newcolumntype{N}{>{\centering\arraybackslash}m{1.5cm}}
\newcolumntype{Q}{>{\centering\arraybackslash}m{1.2cm}}
\newcolumntype{G}{>{\centering\arraybackslash}m{1.2cm}}
\newcolumntype{R}{>{\centering\arraybackslash}m{0.8cm}}
\newcolumntype{P}{>{\arraybackslash}m{2.0cm}}
\newcolumntype{B}{>{\centering\arraybackslash}m{4.2cm}}
\newcolumntype{C}{>{\centering\arraybackslash}m{2.5cm}}
\newcolumntype{D}{>{\centering\arraybackslash}m{0.3cm}}
\newcolumntype{E}{>{\centering\arraybackslash}m{0.7cm}}
\newcolumntype{F}{>{\centering\arraybackslash}m{1.4cm}}
\makeatletter
\def\BState{\State\hskip-\ALG@thistlm}
\makeatother

\usepackage{etoolbox}
\usepackage{tikz}
\usetikzlibrary{tikzmark}
\usetikzlibrary{calc}
\usepackage{xparse}

\NewDocumentCommand{\statcirc}{ O{#2} m }{%
    \begin{tikzpicture}
    \fill[#2] (0,0) circle (1.0ex); 
    \fill[#1] (0,0) -- (180:1.3ex) arc (180:0:1.3ex) -- cycle; 
    \end{tikzpicture}
}
\errorcontextlines\maxdimen

\newcommand{\ALGtikzmarkcolor}{black}
\newcommand{\ALGtikzmarkextraindent}{4pt}
\newcommand{\ALGtikzmarkverticaloffsetstart}{-.5ex}
\newcommand{\ALGtikzmarkverticaloffsetend}{-.5ex}
\makeatletter
\newcounter{ALG@tikzmark@tempcnta}

\newcommand\ALG@tikzmark@start{%
    \global\let\ALG@tikzmark@last\ALG@tikzmark@starttext%
    \expandafter\edef\csname ALG@tikzmark@\theALG@nested\endcsname{\theALG@tikzmark@tempcnta}%
    \tikzmark{ALG@tikzmark@start@\csname ALG@tikzmark@\theALG@nested\endcsname}%
    \addtocounter{ALG@tikzmark@tempcnta}{1}%
}

\def\ALG@tikzmark@starttext{start}
\newcommand\ALG@tikzmark@end{%
    \ifx\ALG@tikzmark@last\ALG@tikzmark@starttext
    \else
        \tikzmark{ALG@tikzmark@end@\csname ALG@tikzmark@\theALG@nested\endcsname}%
        \tikz[overlay,remember picture] \draw[\ALGtikzmarkcolor] let \p{S}=($(pic cs:ALG@tikzmark@start@\csname ALG@tikzmark@\theALG@nested\endcsname)+(\ALGtikzmarkextraindent,\ALGtikzmarkverticaloffsetstart)$), \p{E}=($(pic cs:ALG@tikzmark@end@\csname ALG@tikzmark@\theALG@nested\endcsname)+(\ALGtikzmarkextraindent,\ALGtikzmarkverticaloffsetend)$) in (\x{S},\y{S})--(\x{S},\y{E});%
    \fi
    \gdef\ALG@tikzmark@last{end}%
}

\apptocmd{\ALG@beginblock}{\ALG@tikzmark@start}{}{\errmessage{failed to patch}}
\pretocmd{\ALG@endblock}{\ALG@tikzmark@end}{}{\errmessage{failed to patch}}
\makeatother

\begin{document}
\begin{frontmatter}
\title{ETGuard: Detecting D2D Attacks using Wireless Evil Twins}

\author[mnit]{Vineeta Jain\corref{cor1}}
\ead{2015rcp9051@mnit.ac.in}
\author[mnit]{Vijay Laxmi}
\author[jammu]{Manoj Singh Gaur}
\author[labri]{Mohamed Mosbah}
\address[mnit]{Malaviya National Institute of Technology Jaipur (MNIT Jaipur), India}
\address[jammu]{Indian Institute of Technology Jammu, India}
\address[labri]{Univ. Bordeaux, Bordeaux INP, CNRS, LaBRI, UMR5800, F-33400 Talence, France}

\cortext[cor1]{Corresponding author. Mobile: +91-7692932513}

\begin{abstract}
In this paper, we demonstrate a realistic variant of wireless Evil Twins (ETs) for launching device to device (D2D) attacks over the network, particularly for Android. An ET can be defined as a rogue Access Point (AP) created by hackers to resemble the authentic AP in a network zone. 
The existing attacks that can be launched through ETs include sniffing, Man-in-the-Middle (MITM) attack, etc. However, these attacks affect the devices after their association and transmission of user traffic through an ET. We show an attack where an ET infects an Android device before the relay of network traffic through it, and disappears from the network immediately after inflicting the device. The attack leverages the captive portal facility of wireless networks to launch D2D attack. We configure an ET to launch a malicious component of an already installed app in the device on submission of the portal page. For example, the malicious component can be either a service which opens a port, or sends an SMS to premium number, or exfiltrates sensitive information to malicious server. Thus, the attack may lead to any number of consequences. 
The existing ET detection solutions on APs are incapable of preventing this attack due to two reasons - either they analyse an ET after the relay of user traffic through it, or they can detect this attack only for hardware ETs. 
In this paper, we present an online, incremental, automated, fingerprinting based pre-association detection mechanism named as ETGuard which works as a client-server mechanism in real-time. The fingerprints are constructed from the beacon frames transmitted by the wireless APs periodically to inform client devices of their presence and capabilities in a network. Once detected, ETGuard continuously transmits deauthentication frames to prevent clients from connecting to an ET. ETGuard outperforms the existing state-of-the-art techniques from various perspectives. Our technique does not require any expensive hardware, does not modify any protocols, does not rely on any network specific parameters such as Round Trip Time (RTT), number of hops, etc., can be deployed in a real network, is incremental, and operates passively to detect ETs in real-time. To evaluate the efficiency, we deploy ETGuard in 802.11a/b/g wireless networks. The experiments are conducted using $12$ different attack scenarios where each scenario differs in the source used for introducing an ET. ETGuard effectively detects ETs introduced either through a hardware, software, or mobile hotspot with high accuracy, only one false positive scenario, and no false negatives.
\end{abstract}

\end{frontmatter}

\section{Introduction}
\label{introduction}
With $2$ billion active users around the world and $87.5$\% of the global market share\cite{android}, Android is currently one of the most popular operating systems. Android has become a prime target of hackers and malware writers mainly due to three reasons - firstly, Android holds an enormous user base and relies on them to take safety critical decisions, such as which server certificates to trust and install?, what permissions to be granted to the app?, is the Internet connection safe or not?, etc. Secondly, due to the flexibility offered by Android, developing apps is not a difficult task
nowadays. Recently, 3 million apps have been recorded on Google play store. These novice developers are unaware and unconcerned about security of the user. For instance, app
developers are free to either mix secure and insecure connections in the same app, or not use SSL at all. Certificate checking and hostname verification are optional in Android and depends on developers. This opens doors for attacks such as MITM, SSL stripping, etc. Thirdly, Android makes security assumptions which can induce it to fall as a prey for severe
attacks such as remote exploitation, sniffing, etc. For example, Android recognises a wireless network by its SSID (Service Set Identifier). SSID is an alpha-numeric string for uniquely identifying an AP. If any wireless network spoofs the SSID (creates an ET) and
arrives in the vicinity of the device, Android assumes it to be the genuine one and connects to it directly. It may lead to dangerous consequences such as information loss, financial loss, damage of the device, remote control of the device, etc. 
ETs can exploit the app vulnerabilities of SSL implementation in Android to launch phishing
attacks. ETs can redirect download and installation requests for apps downloaded from websites instead of Google playstore to malicious APK files. Thus,
Android becomes an easy target for attackers.\\
\indent ETs are easy to create and launch in a network. They can be created simply by spoofing SSID. Further, attacker physically situates the ET adjacent to targeted user base for captivating victim's wireless connection by transmitting stronger signal strength, and continuously sending deauthentication frames for legitimate APs to force victims to lose their present wireless connection. 
Thus, users are tend to get connected to the ET.\\
\indent A lot of work has been conducted in the direction of ET detection. We categorise the existing solutions in two categories - post-association and pre-association. The post-association techniques identify an AP as an ET after associating with it. They use various parameters to identify an ET such as ETSniffer\cite{yang2012active} uses \textit{hop count} to identify an ET. CETAD\cite{mustafa2014cetad} recognises an ET by computing its \textit{Internet Service Provider} (ISP) and \textit{RTT}. Song et al.\cite{shetty2007rogue} calculates \textit{inter-packet arrival time}  between hops to discover an ET. WifiHop\cite{monica2011wifihop} relays watermarked packets on associated wireless network before any user traffic. It monitors the path traversed by watermarked packets to detect an ET. The pre-association techniques disclose an AP as an ET before associating with it. Bratus et al.\cite{bratus2008active} creates fingerprints of \textit{802.11 MAC responses} from routers. However, this approach can not detect an ET launched using mobile phones or softwares. Tang et al.\cite{tang2017exploiting} utilises \textit{Received Signal Strength Indicators} (RSSI) to catch an ET. However, this approach is only for stabilised APs, and can not detect an ET from softwares or mobile devices. Unfortunately, no such technique exists which can effectively and efficiently detect a wireless AP as an ET (launched either through a hardware, software, or mobile phone) before associating with it. \\ 
\indent In this paper, we present a practical approach to conduct D2D attack in the network using ETs. D2D communication is defined as the establishment of link between devices using communication channels (wired or wireless)\cite{Mohanan:2017:PIT:3165118}. Here, we consider the channel to be wireless. We define D2D attack as an exploit where one device ($A$) launches a malicious activity on another device ($B$) through the wireless
communication channel. In our case, we consider $A$ to be either an Android device, or a laptop, or a wireless router, and $B$ to be an Android device. We illustrate an attack using the captive portal feature of wireless networks to attack the device
before relay of network traffic through it. A captive portal is an obligatory web interface for authenticating/managing a user before Internet access is granted. It is mainly used for public places such as airports, educational institutes, enterprise networks, etc., to keep a track of connected users or prevent any external user from getting access to a premises network. We create an ET for a network which adopts captive portal for user management. As soon as user launches the portal, ET launches a malicious service present on the device which secretly opens a port for the attacker to steal sensitive information. After infecting user's device, the ET gets disconnected from the network instantaneously. Thus, the ET
can launch an attack without leaving any trace behind. \\ 
\indent This paper presents an automated, online, incremental, fingerprinting based pre-association solution for detecting ETs in real-time for infrastructure based networks. 
The proposed solution follows a client-server architecture where fingerprints are stored on the server, and a client side interface in the form of an Android app is provided to the users. Whenever a user sends a request for scanning the network for ETs, the app communicates to the server, and the server scans the network by comparing the currently available APs fingerprints with the stored fingerprints. 
The app displays SSID, BSSID (Basic Service Set Identifier) and signal strengths of all the available APs in the network. This provides a detailed view of the network to the user, and if any suspicious AP is encountered, user may request the servers to scan the network via the app.\\ 
\indent The contributions of this work are as follows:
\begin{itemize}
\item This paper presents a practical version of D2D attacks through ETs in the network. 
We illustrate an archetype application of the attack and explain the serious consequences of the attack on Android devices.
\item We propose a fingerprinting based pre-association solution for detecting ETs in an infrastructure based network that outperforms the existing ET detection techniques. The proposed detection mechanism is a client-server solution to detect an ET before the association and relay of user traffic through it. Our technique is incremental, does not require any expensive hardware, does not modify any protocols, does not rely on any network specific parameters such as RTT, number of hops, etc., remains online, and operates passively to detect ETs in real-time. 
\item We utilise beacon frames to detect ETs. Since the fields of beacon frames are hardware, driver, security configurations and signal strength dependent, and vary from hardware to software to mobile hotspots, therefore, beacon frame fingerprinting is effective in detecting ETs in a network.
\item We implement our proposed approach through a prototype known as ETGuard. We have profusely evaluated ETGuard by creating $12$ experiment environments with hardware/software/hotspot ETs introduced in them. ETGuard successfully detects ETs with high accuracy, extremely low false positive rate and no false negatives.   
\end{itemize}
\indent The organisation of the paper is as follows: Section \ref{threat} describes the attack scenario and consequences associated with it. Section \ref{preliminaries} explains the beacon frame fields used for fingerprinting, defines D2D attack, and classifies ET attack scenarios. Section \ref{approach} introduces ETGuard, and describes the architecture and methodology of ETGuard. Section \ref{results} highlights the experimental evaluation of ETGuard on real-time attacking scenarios, and comparison of ETGuard with the existing state-of-the-art techniques. Section \ref{related} explains the state-of-the-approaches for ET detection. Section \ref{Conclusion} concludes the paper with future work.
\section{Threat Model and Assumptions}
\label{threat}

\indent ET attacks had been a focus of researchers for more than a decade. However, the research community considered only the networking aspect of ET attacks, i.e., user traffic sniffing or MITM attack. Sniffing and MITM attacks can harm the user only when user associates to the ET, and start generating network traffic through it. We present an attack scenario which causes damage before the generation of any user traffic through the ET. In this section, we explain the attack scenario - Invoking Malicious Component.
\subsection{Invoking Malicious Component}
In this attack, the attacker can launch any activity/service of an app present on the device. Suppose an attacker repackages a benign app (\textit{U}) by including a malicious service (\textit{V}) in it which sends an SMS to premium numbers. This service is not invoked by any component of the app, and hence, the app hoaxes the scanners and bouncers of playstore and other app markets, and gets published.
\subsubsection{Assumptions}
\indent We assume that there is a wireless network named \textit{CSE} in a premises (such as educational institute, industry, airport, etc.) with the following features:
\begin{itemize} 
\item It uses captive portal facility to grant Internet access. 
\item It does not use any passphrase security.
\item The target Android device has \textit{U} installed which contains \textit{V}.
\item The network is neither behind NAT (Network Address Translation), nor protected by firewalls. 
\end{itemize} 
\subsubsection{Attack Scenario}
The steps used to launch the attack are as follows:
\begin{itemize}
\item We introduce an ET of ``CSE" with the captive portal similar to the original one by using Coovachilli\cite{chilli}, hostapd\cite{hostapd} and Easyhotspot\cite{easyhotspot}. Coovachilli serves as an access controller for the ET. It diverts the client to the fake captive portal page by blocking the wireless traffic. Hostapd performs the function of spoofing and broadcasting the beacon frames for the ET. We spoof SSID and BSSID of the ET to make it similar to ``CSE". Easyhotspot is used to customise the captive portal provided by Coovachilli to look exactly similar to that of ``CSE". Hence, the users do not get suspicious of the portal page received on getting connected to an ET of ``CSE".
\item An ET is placed at a location that provides a higher signal strength than the existing genuine AP. We transmit deauthentication frames for the legitimate AP to disconnect the users from it. Thus, the Android users will be connected directly to the fake AP.
\item As soon as the device gets connected to the ET, it gives a notification of ``Sign into the network" on the user device, similar to the genuine one. Thus, the user does not get suspicious and clicks on it.
\item On clicking the notification, the user is redirected to the maliciously spoofed captive portal page wherein users provide their username and password. When they click on the submit button, the intent \textit{V} gets launched on the device. As soon as \textit{V} is launched, the ET disappears from the network. As the attack launches a service, the user remains unaware that anything fishy has happened on his device, and subsequently gets connected to the legitimate AP. Thus, an ET creeps in the network, harms the device, and vanishes without leaving any traces behind. The attack can prove extremely dangerous for the user, if the service opens a port secretly. The malicious service can either execute netstat command, or read /proc/$<$pid$>$/net/tcp file to identify open ports on the device (we assume that the ports are bind to local IP\footnote{IP address stands for Internet Protocol address} address), and further launch attacks such as information theft, malware installation, remote control, etc~\cite{jia2017open}. Notably, no special permissions are required to scan the ports on the device~\cite{jia2017open}. Thus, the impact of this attack can be dangerous for the user. 
\item Additionally, we steal the username and password entered by the user. This can be misused to get illegitimate access to the network. 
\end{itemize}
\lstset{language=Java, escapeinside={(*@}{@*)}, basicstyle=\ttfamily\scriptsize}    
\begin{lstlisting}[caption= Code snippet for launching an activity or service of an app on Android device using ET]
	<a href="intent://scan/#Intent;
			scheme=my.special.scheme.V;
	package=com.example.vineeta.
			U;end">
	<input name="accept" type="button" 
		value="LOGIN" style="cursor:pointer;"></a>
\end{lstlisting}

\indent Listing $1$ displays the code for launching the service $V$ of the app $U$ installed on the device using the fake captive portal page of an ET. In Listing 1, the tag \textit{$<$input name=``accept" type=``button"
value=``LOGIN" style=``cursor:pointer;"$>$} denotes the HTML (Hypertext Markup Language) code of the LOGIN button on the portal page. The HTML tag \textit{$<$a$>$} preceding the code of the LOGIN button explains that a link opens on clicking that button. However, in place of the link, the attacker maliciously invokes a service by launching an intent. Intents are used in Android for inter-component communication among apps~\cite{intent}. The components of an app can summon other components of the same or different apps using Intents. The code \textit{intent://scan/\#Intent;scheme=my.special.scheme.V; package=com.
example.vineeta.U;}  shows that the intent will invoke the service $V$ of the app $U$ (containing package name as com.example.vineeta.U) supporting the scheme my.special.scheme.V. Package name uniquely identifies an app on the device, and scheme defines the type of the data handled by the intent~\cite{intent}. Scheme can be either pre-defined (such as images, videos, etc.), or custom-defined (such as mail)~\cite{intent}. Thus, as the user clicks on the LOGIN button of the fake portal page, the malicious service gets launched on his device. 
\begin{figure}[htbp]
    \centering
    \includegraphics[scale=0.3]{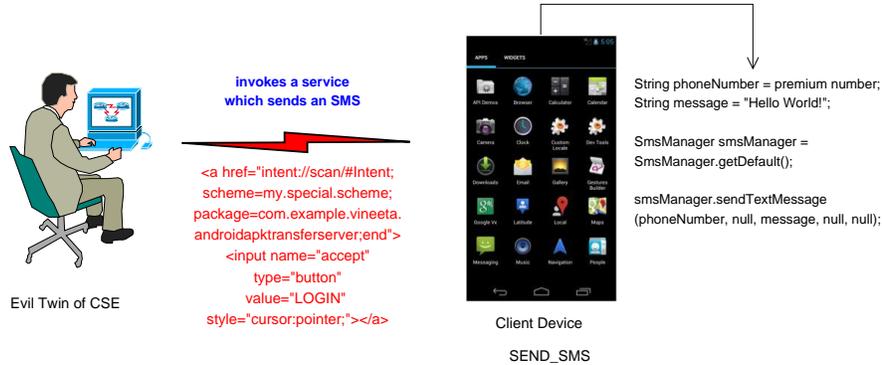}
    \caption{Invoking the malicious service which sends an SMS to premium numbers.}
    \label{V}
\end{figure}
\\
Figure \ref{V} describes the attack scenario where the invoked service is sending messages to premium numbers. Thus, the D2D attack using ETs may have deadly ramifications.
\\
\indent Since captive portal is a web page which authenticates and accounts the activities of the user in a network, many web based attacks can be launched through the portal. Xia et al.~\cite{xia2004detecting} have demonstrated the launch of session hijacking attack through captive portal. In this attack, the attacker transmits deauthentication frames to disconnect the genuine user from the network, and mimics the MAC (Media Access Control) address and DHCP (Dynamic Host Configuration Protocol) network configuration of the user. Thus, the attacker acts as a valid user in the network by hijacking the session. Adrian et al.~\cite{dabrowski2016browser} launch browser history stealing attack through captive portal. The attacker modifies the portal page by including a large number of image references with links. The page checks the links against the persistent cookies database of the user's device, and logs the cookies which are associated to the links in the database. However, this attack requires a huge DNS (Domain name System) lookup time (which can make a user suspicious), and a large database of websites against which the cookies of the users are checked. Chen et al.~\cite{chen2010proof} proposes a MITM attack using ARP (Address Resolution Protocol) spoofing to bypass the captive portal, and access Internet without getting authenticated. The attacker locates itself between authenticated client and AP, and conduct IP masquerading to access Internet. \\
\indent A lot of ET attacks have also been proposed in the literature. Karma attack~\cite{karma} sniffs the probe request of the client, and on the basis of requested parameters, automatically creates an AP on the fly. Catch-All-Evil-Twin attack~\cite{catch} creates multiple ETs for a single SSID with all types of security configurations. DNS Spoofing~\cite{sax2000dns} and SSL Stripping~\cite{moixe2009new} attacks are launched after acquiring a MITM position between client and AP in the network. In DNS Spoofing, the attackers replace the requests from clients with their choice of web pages. In SSL Stripping, attackers convert HTTPS (Secured Hypertext Transfer Protocol) link to HTTP link on the fly. \\   
\indent The above mentioned attacks either cause loss of information to user, or remove the user from the network, or perform the classic attacks (such as sniffing, MITM attack) in a different way. But, neither of these attacks can harm the user before getting connected to Internet, nor harm the device of the user, nor cause financial loss directly. \textit{Invoking Malicious Component} attack introduces an ET in the network, fools an Android user by presenting a cloned captive portal, and when the user clicks on the LOGIN button, invokes an already present malicious component of an app. This component may cause financial loss (such as sending an SMS to premium number), or information loss (such as sharing the location, browser history), or provide a backdoor to the device (such as open a port). Notably, all attack steps are performed before the relay of network traffic, and the attacker needs not to be present in the network for a long time. As soon as the device gets infected, attacker can simply switch the ET off, and allow the user to connect to the genuine AP. So, tracking the ET is extremely challenging. Therefore, a technique is required to scan for ETs in a network before the establishment of connection between the ET and the device.
 




\section{Preliminaries}
\label{preliminaries}
In this section, we explain beacon frame components of APs which contribute in ET detection, define D2D attacks on Android, and classify ETs and ET attack scenarios in three categories on the basis of ecosystem and the position of the ET in the network, respectively.\footnote{The paper follows this classification and terminology.}
\subsection{Beacon Frame Components}
\label{ieee}
Beacon frame is a management frame transmitted by an AP for broadcasting its capabilities to client devices in a network. 
Beacon frame body encloses information about the properties of the AP. It includes mandatory and optional fields of fixed and variable lengths. The variable length fields are known as Information Elements (IEs)\cite{mgmt}. In this paper, we are employing beacon frame fingerprinting to differentiate between a genuine AP and its ET. The following fields are used for fingerprinting:
\begin{itemize}
\item \textit{Beacon Interval}: It illustrates the time between two beacon frame transmissions. The default value is $102.4$ milliseconds.
\item \textit{Capability Information}: It contains 14 subfields which represents the required network capabilities which should be fulfilled by a client station, in order to connect to an AP. For example, if the value of Wireless Equivalent Privacy (WEP) flag of an AP is 1, only those clients which also enforces WEP for privacy, are permitted to connect to the AP.
\item \textit{SSID}: It is an alphanumeric string which uniquely recognises an AP in a network. The APs are visible to user by these names. It is a variable length IE whose length lies in the range of $0-32$ bytes. 
\item \textit{Supported Rates}: It is a variable length IE that displays all the mandatory and supported data rates by an AP. An AP must have at least one mandatory rate, and can have multiple supported rates. Any client station that wishes to get connected to an AP must support all mandatory data rates. 


\item \textit{Traffic Indication Map (TIM)}: 
TIM field of beacon frame is a variable length IE that represents information about the sleeping client stations with pending frames, and intervals during which the AP attempts to deliver the frames. 
Every beacon frame does not carry information about buffered frames. An element known as Delivery Traffic Indication Map (DTIM) carries information about the interval between beacon frames before a beacon containing details of buffered frames is transmitted. It varies from AP to AP.   
\item \textit{Country}: There are regulatory bodies in every country that impose restrictions in their domains regarding the permitted channels and power levels. It is a variable length IE which specifies the country whose restrictions the AP follows, the permitted channels and the maximum power levels followed by the AP.  
\item \textit{Robust Security Network (RSN)}: An AP can administer encryption techniques for unicast and multicast traffic in a network. RSN is a variable length IE that demonstrates details about cipher suites used for authentication and encryption. It also defines other RSN capabilities such as authentication key management.   
\item \textit{Extended Support Rates}: the beacon frame field known as Supported Rates can only store values for $8$ data rates. If any AP wishes to support more than $8$ data rates, that information is represented in Extended Support Rates IE. 
\item \textit{Vendor-Specific}: It is a variable length IE (up to 252 bytes) which is provisioned to always be present as a last element in the frame body of beacon. This is common for every OEM (Original Equipment Manufacturer) to put their own proprietary information within the beacon frames.

\item The beacon frame also consists of a MAC header. MAC stands for Media Access Control. It is a sub-layer of DLL (Data link layer) in OSI (Open System Interconnect) model. All the components of 802.11 falls under the MAC layer of DLL. All the fields in the MAC header of management frame are mandatory and of fixed length. From the MAC header fields, we have taken in use \textbf{BSSID} field as it signifies the MAC address of the AP to uniquely identify it in a network. 
\end{itemize}
\subsection{Definition of D2D Attack}
\label{definition}
``D2D communication is defined as the establishment of link between devices using communication channels (wired or wireless). 
D2D attack is defined as an exploit where one device ($A$) launches a malicious activity on another device ($B$) through the wireless communication channel"\cite{Mohanan:2017:PIT:3165118}. In our case, we consider $A$ to be either an Android device, or a laptop, or a wireless router, and $B$ to be an Android device.
\subsection{Types of ETs}
This classification is based on the type of environment used for creating ETs. They can be created using the following three methods:
\begin{itemize}
\item \textbf{Hardware}: A hardware device such as router can be used to introduce the ET in a network. This is an expensive method, and thus, rarely used by the attackers. 
\item \textbf{Software}: This is the most prevalent method for introducing ETs in a network. Many open source softwares are available to create ETs using laptops such as hostapd, Connectify, ap-hotspot, aircrack-ng, etc.
\item \textbf{Mobile devices}: ETs can be created using tethering facility of smartphones. This is apparently easy than the other methods as it does not require any additional expertise. However, they cater the attackers comparatively less control over the ET, as they allow the attackers to modify very limited number of parameters such as SSID and BSSID. Thus, they are not preferred for specialised ET attacks. 
\end{itemize} 
\subsection{Launching ET Attacks}
\label{launching}
We have classified ET attack scenarios in a network in three classes on the basis of location of an ET in the network:
\begin{figure*}[htbp]
\centering
\includegraphics[scale=0.3]{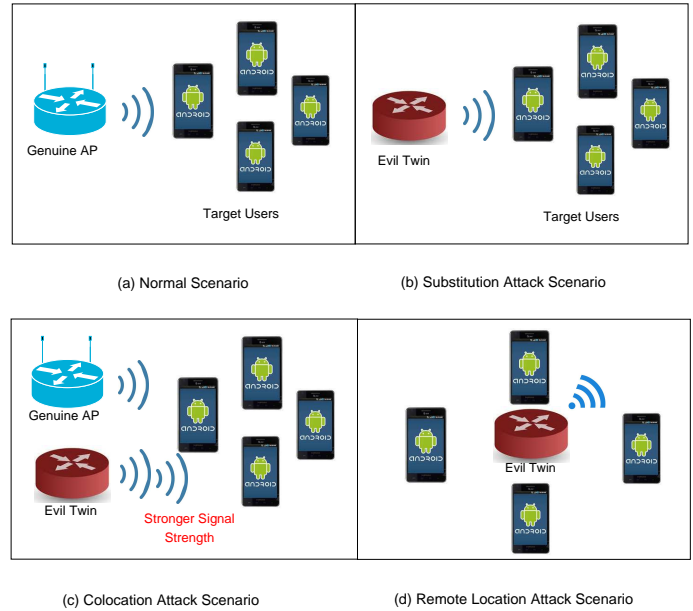}
\caption{Classification of attack scenarios on the basis of position of ET in the network.}
\label{scenario}
\end{figure*}
\begin{itemize}
\item \textbf{Substitution}: In this scenario, attacker replaces the genuine AP with the ET at the same location. 
\item \textbf{Colocation}: In this scenario, attacker colocates the ET near the genuine AP, and transmits stronger signal strength to allure users to connect to them. 
\item \textbf{Remote Location}: In this scenario, the genuine AP was previously present in the network. Thus, the profile of genuine AP exists in the devices of users. When the attackers introduce ET in this network at any location, users automatically get connected to the ET, due to the saved profile in the devices of users. This scenario is different from Substitution attack scenario in terms of location of the ET, and Colocation attack scenario in terms of the signal strength. In this case, ETs need not to transmit stronger signal strength, and the location of the ETs in the network need not to be same as the genuine AP. 
\end{itemize}
Figure \ref{scenario} describes the three attack scenarios used for launching ET attacks.
\section{ETGuard}
\label{approach}
This section explains the workflow and methodology of ETGuard to detect ETs in an infrastructure based network.
\subsection{ETGuard Overview} 
We present ETGuard, an online, automated, incremental, real-time pre-association analysis tool that accumulates the unique fingerprints of legitimate APs present in a network, and provides a client interface in the form of an Android app to users for effectively scanning the network to detect ETs in real-time. Figure \ref{fig:1a} shows the workflow of ETGuard. It consists of the following $7$ modules:
\begin{figure*}[htbp]
\centering
\includegraphics[width=\textwidth]{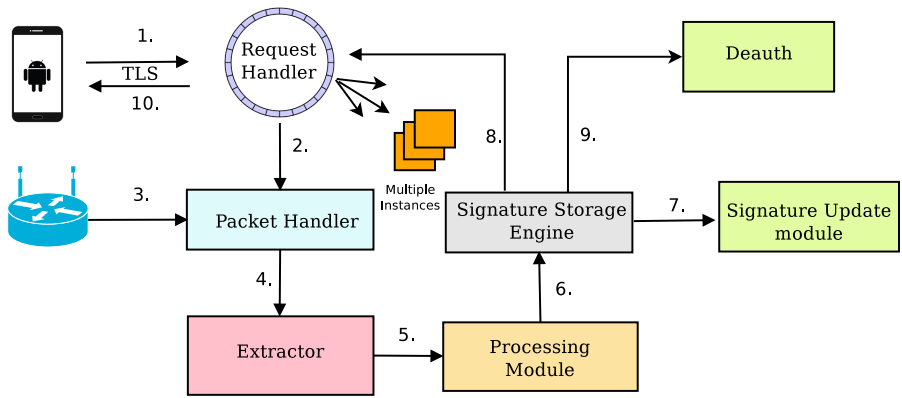}
\caption{Workflow of ETGuard}
\label{fig:1a}
\end{figure*}
\begin{itemize}
\item \textbf{Request Handler}: It concurrently captures the ET detection requests from devices, and forwards it further to Packet Handler module. It is implemented using a circular buffer, and can create multiple instances of itself when the user requests reach the threshold. Thus, it is capable of preventing denial of service attacks. The transmission of requests from user device to the Request Handler and vice versa are protected by TLS (Transmission Layer Security) to prevent tampering and eavesdropping. 
\item \textbf{Packet Handler}: The Packet Handler module collects beacon frames of APs present in the vicinity.   
\item \textbf{Extractor}: The Packet Handler successively transmits the captured beacon frames to the Extractor. The Extractor module extracts the relevant fields from the beacon frames to construct fingerprints. 
\item \textbf{Processing Module}:
The extracted fields are merged to construct fingerprints for the APs. If any field is not present in an AP, it is substituted by NULL. Further, these fingerprints are transferred to Fingerprint Storage Engine. 
\item \textbf{Fingerprint Storage Engine}: The fingerprints are matched across the stored fingerprints. If a perfect match is found, the AP is a genuine AP. The partial matching rules and ET detection is explained in detail in Subsection \ref{detect}.
\item \textbf{Fingerprint Update Engine}: If any new AP is found in the vicinity, it gets added in the fingerprint database. Thus, the system is incremental. Even if any of the beacon frame fields of any genuine AP gets modified, the corresponding fingerprint is updated by this module. 
\item \textbf{Deauth}: This module broadcasts deauthentication frames for the detected ETs to forcefully disconnect all its associated clients.
\end{itemize}
\begin{figure*}[htbp]
\centering
\includegraphics[scale=0.3]{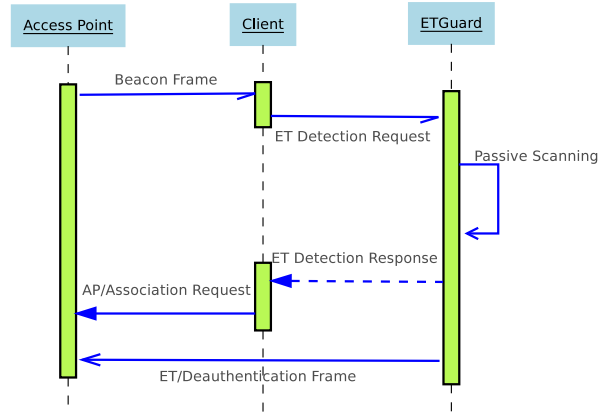}
\caption{Sequence Diagram of ETGuard}
\label{fig:1b}
\end{figure*}
\indent The modules of ETGuard are sequenced in time. The architecture of ETGuard consists of two components - \textbf{server} and \textbf{client}. The client side component includes the app interface for users, and server side component handles the beacon frame fingerprinting for a network.
Figure \ref{fig:1b} shows the sequence diagram of ETGuard in the network. Before associating with any AP, client device sends ET detection request signals to the ETGuard server through the ETGuard client app to passively scan the network for ETs\footnote{In the rest of paper, server stands for ETGuard server and client stands for ETGuard client app}. Further, server captures the beacon frames, and matches the fingerprints with the legitimate fingerprints to analyse the genuineness of the AP. The server responds back with an ET detection response, and subsequently if the AP is not an ET, client device sends association request to the AP. Otherwise, the app highlights the ET with red color in the WI-FI list for the users. ETGuard server also broadcasts deauthentication frames for the ET to forcefully disconnect all its associated clients. Thus, it prevents user from associating to ETs in the network. \\
 
\subsection{Server Analysis of ETGuard}
\label{serverana}
In order to detect ETs in a network, we incorporate beacon frame fingerprinting mechanism. The server accommodates the beacon frame fingerprints of legitimate APs. For sniffing ETs, server passively scans the network by capturing the beacon frames on all available channels, and compares it with the stored fingerprints (complete algorithm is explained in Subsection \ref{detect}). Before we explain the functionality of server, we need to answer a few questions:
\begin{itemize}
\item Do all fields of beacon frame contribute in ET detection?
\item Why beacon frame fingerprinting is efficient in differentiating legitimate AP from the fake ones?
\item Do the fields that change dynamically in beacon frames in a real environment can contribute in ET detection?
\item If  AP and ET belong to the same OEM, can ETGuard detect the fake one?
\end{itemize}
\indent \textit{For the first question}, all the fields in beacon frame do not contribute in fingerprinting mechanism of ETGuard, as some of the fields differ from beacon to beacon of an AP in a real network scenario, such as channel information, timestamp values, sequence number, frame check sequence, etc. Only those fields whose values remain constant across all the beacon frames transmitted by an AP are considered for fingerprinting. The fields used for beacon frame fingerprinting are: \texttt{BSSID, Beacon Interval, Capability Information, SSID, Supported Rates, TIM, DTIM, Country, RSN, Extended Support Rates, and Vendor-Specific}. Out of the above 11 fields, 6 fields are optional in beacon frame. To judge whether these $6$ optional fields of beacon frames are present in the APs of real world or not, we captured the beacon frames of about $50$ APs, and created a dataset named \textbf{MNIT} dataset. Since the dataset is prepared by capturing beacon frames of APs from public places (such as railway stations, airports), educational institutions, coffee shops and corporate organisations, therefore, the dataset gives a fair knowledge about the presence of these fields in practical scenarios. By analysing the dataset, we observed that these $6$ optional fields are present in $95\%$ of the APs. Table \ref{dataset} presents the experimental results of 10 APs from the dataset.\\
\begin{table}[htbp]
\begin{center}
\caption{Analysis results of MNIT Dataset}
\label{dataset}
\tiny{
\begin{tabular}{|c|c|c|c|c|c|}
\hline
\rule{0pt}{0.5cm} 
\textbf{SSID of the AP*}	& \textbf{TIM} & \textbf{Country} & \textbf{Extended Support Rates} & \textbf{RSN} & \textbf{Vendor-Specific} \\[1ex]
\hline
SSID 1 & \cmark & \cmark & \cmark & \xmark & \cmark \\[1ex]
\hline
SSID 2 & \cmark & \cmark & \cmark & \cmark & \cmark\\[1ex]
\hline
SSID 3 & \cmark & \cmark & \cmark & \cmark & \cmark \\[1ex]
\hline 
SSID 4 & \cmark & \cmark & \cmark & \cmark & \cmark   \\[1ex]
\hline
SSID 5 & \cmark & \cmark & \cmark & \cmark & \cmark \\[1ex]
\hline
SSID 6 &\cmark & \cmark & \cmark & \cmark & \cmark \\[1ex]
\hline
SSID 7 & \cmark & \xmark & \cmark & \cmark & \cmark \\[1ex]
\hline
SSID 8 & \cmark & \xmark & \cmark & \cmark & \cmark \\[1ex]
\hline
SSID 9 & \cmark & \cmark & \cmark & \cmark & \cmark \\[1ex]
\hline
SSID 10 & \cmark & \cmark & \cmark & \cmark & \cmark \\[1ex]
\hline
\end{tabular}}
\end{center}
*We have anonymised the names of APs for security purposes.
\end{table}
\indent \textit{For the second question}, beacon frame is a frame transmitted by an AP to broadcast its capabilities to clients in a network. The capabilities differ from OEM to OEM for hardware AP, from hardware AP to software AP, from hardware AP to mobile hotspot, and from mobile hotspot to software AP. Following are the arguments for using beacon frame fingerprinting:
\begin{itemize}
\item \texttt{TIM} element in the beacon frames is used for buffering data for low-power devices. This is a non-configurable field as it depends on the load of buffered data. We observed that the default TIM element length (when no data is buffered) for beacon frames of mobile hotspot is different than the hardware and software APs. For mobile hotspot the default length is $9$, whereas for software and hardware APs, the default length is $4$. Since, it is not a configurable field in beacon frames, thus, the difference in the lengths of this element in the beacon frame aids in the detection of ET through mobile hotspots.
\item \texttt{Capability Information} element highlights the various capabilities of an AP such as whether an AP can support WEP or not, whether an AP is using block ACK or not, etc. Some of the capabilities are hardware and driver dependent, and could not be modified such as delayed acknowledgement, immediate acknowledgement, use of short preamble, etc. If an ET is created using a software AP or different hardware, capabilities differ. 
\item \texttt{Supported Rates} element holds different rates for different APs. A hardware AP supports more rates as compared to software AP and hotspot. APs also contain an Extended Support Rates element which contains additional rates supported by an AP. By default, software APs contain less mandatory fields and more Extended Support Rates. However, in hardware APs, extended ones are less and mandatory ones are more. This field is also hardware and driver dependent. 
\item \texttt{Vendor-Specific} element is different for different OEMs, and present in all hardware APs. It cannot be modified as it contains proprietary information about vendors. 
\item An attacker can possibly use social engineering to identify the security passphrase for the legitimate AP. If an attacker adopts the similar RSN settings and BSSID as of legitimate one, this will prevent user from directly getting connected to the ET\cite{same}, as two APs with same BSSIDs and security settings on a channel restricts Android device to instantly connect to an AP. Thus, for alluring users to promptly connect to an ET, attacker keeps last digit of BSSID different from the legitimate one.
\end{itemize}
Hence, beacon frame fingerprinting is an effective technique to recognise an ET in an infrastructure based network.\\
\indent \textit{For the third question}, the dynamically changing fields such as timestamp, sequence number, TIM, etc., represent the network parameters and load of an AP. We use TIM Information Element for ET detection. 
For the TIM element, DTIM count (explained in Subsection \ref{ieee}) remains constant, and used for fingerprinting. The value of DTIM count differs from OEM to OEM and hardware to software AP. For hardware APs the value is 1, and for software APs and mobile hotspots, the value is 2. The hardware AP and mobile hotspot does not provide user control to change the field. In case of software AP, hostapd provides user control for configuration of this field in beacon frames.\\  
\indent \textit{For the fourth question}, the default configuration fields such as Country, Supported Rates, Extended Support Rates, Vendor-Specific, are similar in APs belonging to same OEM. Thus, if an attacker manages to introduce a hardware ET of the same OEM in the network by forging its SSID and BSSID, and the genuine AP is not using any type of security, beacon frame fingerprinting is not sufficient for detection of ETs. For this purpose, we enforce signal strength indicator fingerprinting in ETGuard to detect ETs belonging to same OEM. The transmitted beacon frame also carries a header named \textbf{Radiotap Header}. It includes information about the signal strength of the AP. The signal strength of two APs can never be same, and cannot be forged. Two APs located at two different places will always transmit different signal strengths. However, signal strengths may oscillate due to hazy effects of radio signals\cite{radio}. But it always lies in a fixed range. In order to attract users, ETs have to transmit a signal strength higher than the highest signal strength offered by the legitimate AP. Hence, ETGuard also stores the highest signal strength offered by an AP to justify its legitimacy in case of ET and AP belonging to the same OEM. 
\subsection{Methodology}
This subsection algorithmically explains the methodology adopted at the server side of ETGuard for detection of ETs.\\
\indent Suppose, there is an infrastructure network I$_n$ with a legitimate AP named as $P$. Let the beacon frame of $P$ be denoted as $P_B$. ETGuard fingerprints the hex values of the following fields of $P_B$:
\begin{itemize}
\item $P_{SSID}$: SSID 
\item $P_{BSSID}$: BSSID
\item $P_{BI}$: Beacon Interval
\item $P_{CI}$: Capability Information
\item $P_{SPR}$: Supported Rates
\item $P_{DTIM}$: DTIM Period
\item $P_{TIM}$: TIM Length
\item $P_{CON}$: Country
\item $P_{ESR}$: Extended Support Rates
\item $P_{RSN}$: Robust Security Network
\item $P_{VEN}$: Vendor-specific
\end{itemize}
ETGuard also stores the highest signal strength offered by the AP separately as $P_{SSI}$\footnote{SSI stands for Signal Strength Indicator}. Let the fingerprint of $P$ be represented as $P_F$. The fingerprint of the AP is constructed as:
\begin{dmath}
\label{e3}
P_F = P_{SSID}\ \Vert\ P_{BSSID}\ \Vert\ P_{BI}\ \Vert\ P_{CI}\ \Vert\ P_{SPR}\ \Vert\\ P_{DTIM}\ \Vert\ P_{TIM}\ \Vert\ P_{CON}\ \Vert\ P_{ESR}\ \Vert\ P_{RSN}\ \Vert\ P_{VEN}
\end{dmath}
For the ETs constructed using software and mobile phones, the fingerprints will be different, and thus, will get detected. Let us discuss a case where an attacker can evade this detection mechanism. Suppose, an attacker introduces a hardware ET of $P$ by using the device of exactly same model and OEM as of $P$. Let us suppose, a perfect match is found between the fingerprints of $P$ and $P'$ as devices of same model and OEM have exactly same hardware and driver values. It then compares the signal strength of the $P$ with the stored value of $P_{SSI}$. Since ETs attract users by offering higher signal strength than the genuine ones, therefore, the ET gets detected.
\subsection{Detection Algorithm}
\label{detect}
Whenever a user requests ETGuard to analyse network for ETs, ETGuard momentarily captures beacon frames and extracts the hex values of various fields. The Radiotap Header (containing information about the SSI field) and header of beacon frame are the first two hex chunks in any beacon frame. They are of fixed length. The fields of beacon frame body are variable in length, and furthermore, they can be present in any order in beacon frame. Therefore, we utilise the IEEE 802.11 unique numerical identifiers for extracting the hex values of these fields. Table \ref{iden} displays the identifiers of fields used for detection in ETGuard. 
\begin{table}[htbp]
\caption{Identifiers for fields of beacon frame body}
\label{iden}
\begin{center}
\footnotesize{
\begin{tabular}{ |c|c|c| } 
 \hline
 \textbf{S.No}. & \textbf{Frame Body Field} & \textit{Identifier} \\ 
 \hline
 1 & SSID & 0 \\ 
 \hline
 2 & Supported Rates & 1 \\ 
 \hline
 3 & TIM & 5 \\ 
 \hline
 4 & Extended Support Rates & 50 \\ 
 \hline
 5 & Vendor-Specific & 221 \\ 
 \hline
 6 & RSN & 48 \\ 
 \hline
 7 & Country & 7 \\ 
 \hline
\end{tabular}}
\end{center} 
\end{table}
After extracting the hex values, Algorithm \ref{hamming} is applied to detect whether the AP is an ET or not.\\ \indent ETGuard classifies the AP in three categories:
\begin{itemize}
\item Legitimate AP
\item an ET
\item Unregistered on the network 
\end{itemize}
If the extracted fingerprint does not match with any of the stored fingerprints, ETGuard compares the SSID to confirm intention of the AP. If the SSID matches with the genuine one, ETGuard identifies it as an ET, else an unregistered AP. \textbf{Thus, ETGuard is capable of detecting any and every type of ET in an infrastructure based network.}
\begin{algorithm}
\caption{Detection Algorithm}
\footnotesize
\label{hamming}
\begin{algorithmic}[1]
\Procedure{ETGuard($P'_B$)}{}
\State $P'_F \gets \textit{Fingerprint of P'}$
\State $D_F \gets \textit{Database of genuine fingerprints}$
\For {\textbf{each} $P_F$ in $D_F$}
\If {$P'_F\neq P_F$}
\If {$P'_{SSID}\neq P_{SSID}$}
\State $\textit{P' is an unregistered AP}$
\Else  
\State $\textit{P' is an ET of P}$
\EndIf
\Else
\State $P'_{SSI} \gets \textit{Signal Strength of P'}$
\If {$P'_{SSI} > P_{SSI} $}
\State $\textit{P' is an ET of P}$
\Else
\State $P \equiv P'$
\EndIf
\EndIf
\EndFor
\EndProcedure
\end{algorithmic}
\end{algorithm}

\section{Results and Discussion}
\label{results}
We implemented the detection algorithm through a prototype named ETGuard (Evil Twin Guard) to evaluate usability, accuracy and performance of the proposed approach. This section explains the implementation of ETGuard, and describes a case study conducted in our institute to evaluate the efficiency of ETGuard in a real network scenario.
\subsection{Implementation} 
\label{implementation}
\indent ETGuard consists of $7$ modules (as explained in Section \ref{approach}). The Request Handler collects the request of users, and concurrently forwards it to Packet Handler. The Packet Handler captures beacon frames using \texttt{tshark}\cite{tshark} on $13$ channels of the network in pcap\cite{pcap} format. Further, the pcap files for all the channels are merged in a single file which is provided as an input to the Extractor. It extracts the relevant fields from the beacon frames one by one, and forwards it to the Processing Module. The pcap file contains redundant beacon frames as an AP transmit a beacon frame every $102.4ms$\cite{mgmt}. To avert the extraneous time and resources in analysing an AP twice, Extractor uses the \texttt{sequence Number} field present in the header of beacon frame. It stores the sequence number of the analysed beacon frame, and if the subsequent beacon frame contains a sequence number one greater than the previous one, it does not analyse the frame, and updates the sequence number with the new one. This avoids the repetitive fingerprint creation of the APs. The Processing Module constructs the fingerprints, and Fingerprint Storage Engine further checks it against the mysql fingerprint database (according to the Algorithm \ref{hamming}). The analysis results are generated in real-time as query processing in mysql database is extremely fast. 
\subsection{Case Study}
We conducted a case study in MNIT Jaipur\cite{mnit} campus network environment to assess the launch of D2D attacks on Android devices, before and after the implementation of ETGuard in the network. 
\subsubsection{Dataset}
We created a fingerprint database for legitimate APs present in the MNIT Jaipur campus using \texttt{mysql} database. The beacon frames are captured using tshark, hex values are extracted using python scripts, and relevant field values are stored in mysql database named \texttt{eviltwin}. We have automated the process of fingerprint generation to remove the manual errors. We have also implemented a Fingerprint Update Module to update the fingerprints and add a new one. Thus, the fingerprinted database is incremental, and can be extended upto a large number of APs.
\subsubsection{Experimental Setup}
\label{experiment}
We set up ETGuard under MNIT Jaipur campus network environment on a laptop (ubuntu 14.04) with a wireless network card. To enable client-side detection, we installed the ETGuard client app on the Android devices of the scholars in phd lab. The app is installed on $5$ different versions of Android (Jelly Bean\cite{jelly}, Kitkat\cite{kitkat}, Lollipop\cite{lollipop}, Marshmallow\cite{marshmallow} and Nougat\cite{nougat}), and devices of $4$ different OEMs such as Lenovo, Sony, Redmi, and Motorola. We define a \textbf{normal scenario} as an environment where there are legitimate AP \texttt{CSE}, ETGuard and users present in the network. We define an \textbf{attack scenario} as the environment setup where legitimate AP \texttt{CSE}, ETGuard, users and an ET of \texttt{CSE} are present in the network. In normal scenario, devices are connected to the legitimate AP, and ETGuard server is waiting for detection requests. In attack scenario, an ET of \texttt{CSE} gets introduced in the network. 
\subsubsection{Detection}
To evaluate the efficacy of ETGuard in detecting ETs in a network, we launch ETs using hardware, software and mobile hotspot in an environment where ETGuard is already present. We analyse the efficiency of ETGuard on the channels supporting IEEE 802.11a/b/g protocols.\\
\indent The hardware ETs used for the experiments include D-Link DIR-615 Wireless-N300 Router\cite{dlink}, Digisol DG-HR1400 Wireless Broadband Router\cite{digisol}, TP-Link TL-WR841N 300Mbps Wireless-N Router\cite{tplink} and Mi 3C Router\cite{mi}. We configure the SSID, BSSID, channel and frequency band of these hardware ETs similar to the legitimate AP \texttt{CSE}. The signal strength of the ETs is kept higher than the legitimate ones to attract user traffic. ETGuard successfully detects all the hardware ETs. Table \ref{comparisontabletechniques} enumerates the similar and disjoint fields in the beacon frames of ETs as compared to the genuine one. Table \ref{comparisontabletechniques} clearly highlights that many parameters of these ETs differ from the legitimate one. The Digisol DG-HR1400 Wireless Broadband Router and D-Link DIR-615 Wireless-N300 Router carry different Capability Information, Supported Rates, Extended Support Rates and Vendor-Specific information than the legitimate one. The TP-Link TL-WR841N 300Mbps Wireless-N Router has similar Capability Information field, but contains distinct Supported Rates, Extended Support Rates and Vendor-Specific fields. The Mi 3C Router carries an additional Country field in the beacon frame which is not present in the legitimate one. Even if the original AP does not contains RSN settings, ETs will be detected due to the difference in other fields. Thus, \textbf{ETGuard can effectively detect the ETs created using hardware of different OEMs.}\\
\indent We introduce an ET named \texttt{CSE} using the similar hardware as of the legitimate one\footnote{We are not revealing the OEM of AP due to security reasons}. Thus, the beacon frame values other than BSSID, RSN and SSI are similar by default. We forge the BSSID, do not implement any type of security, and maintain a higher signal strength. As the original AP is using RSN, ETGuard detects the ET. Further, we conduct an experiment with similar security settings in the ET as of the original one. As mentioned before in Subsection \ref{serverana}, if security settings are applied, we need to modify the BSSID for launching attack on the same channel. We modify the last digit of BSSID for performing the attack. Hence, ETGuard identifies the ET using the BSSID field. Next, we maintain the values of RSN and BSSID similar to the legitimate one. However, we are transmitting a greater signal strength for attracting users, and therefore, the ET gets detected by using the SSI field. Hence, \textbf{ETGuard is capable of detecting ETs belonging to the same hardware.} \\
\indent We introduce the software AP named \texttt{CSE} with the captive portal similar to the original one by using Coovachilli\cite{chilli} + hostapd\cite{hostapd} + Radius Server\cite{radius} + Easyhotspot\cite{easyhotspot}. Coovachilli serves as an access controller for the hotspot. It diverts the client to the captive portal page by blocking the wireless traffic. Hostapd performs the function of configuring and broadcasting the beacon frames for the wireless network. Radius server is used to maintain the authentication and authorisation in the wireless networks. Easyhotspot is used to customise the captive portal provided by Coovachilli. We configure the beacon frame fields of the ET similar to that of legitimate one. However, some of the fields could not be modified, such as in the Country element, the maximum transit power level option can not be set manually, Vendor-Specific element can not be forged, etc. Due to the difference in above fields, ETGuard detects the presence of ET in the network. \\
\indent Additionally, we launched ETs using other open source softwares, even if they don't permit the captive portal facility. We created ET of \texttt{CSE} using ap-hotspot\cite{aphotspot}, aircrack-ng\cite{aircrack} suite and default unity network manager\cite{network} of ubuntu OS (Operating System). Table \ref{os} specifies the operating system and wireless chipset details that are used for launching ETs in the network. 
\begin{table}[htbp]
\caption{Operating systems and drivers used in experiments of software access points.}
\label{os}
\footnotesize
\begin{center}
\begin{tabular}{ |p{1cm}|p{3cm}|p{4cm}| } 
 \hline
 \textbf{S.No}. & \textbf{Operating System} & \textbf{Wireless Chipset} \\ 
 \hline
 1 & Ubuntu 14.04 LTS & Intel Corporation Wireless 3160 (rev 93) \\ 
 \hline
 2 & Ubuntu 16.04 &  Qualcomm Atheros QCA9565 / AR9565 Wireless Network Adapter (rev 01) \\ 
 \hline
 3 & Ubuntu 12.04 & Broadcom Corporation BCM4312 802.11b/g LP-PHY (rev 01) \\ 
 \hline
\end{tabular}
\end{center} 
\end{table}
These softwares does not provide a good amount of user control. The configurable fields permitted by these softwares are modified, but still many of the fields differ. Table \ref{comparisontabletechniques} shows the similar and dissimilar fields of these softwares as compared to the legitimate AP \texttt{CSE}. We can observe from the table that aircrack-ng does not contain many beacon frame fields such as TIM, DTIM, Extended Support Rates and Vendor-Specific. It does not provide user control to modify these fields. One interesting thing to notice is that, DTIM Period of software APs is by default $2$. We successfully detected ETs created by the above softwares.\\
 \renewcommand{\tabcolsep}{0.2cm}
 \renewcommand{\arraystretch}{0.4}
 \begin{sidewaystable*}[htbp]
	\centering
	\tiny
\begin{tabular}{|R|P|Q|G|G|Q|Q|G|G|G|Q|Q|Q|Q|}
 \toprule [0.12em]
   & \tiny{Evil Twin} & \tiny{Beacon Frame Length}& \tiny{SSID}& \tiny{BSSID}& \tiny{Beacon Interval}& \tiny{Capability Information}
   & \tiny{Supported Rates} & \tiny{TIM Length} & \tiny{DTIM Period} & \tiny{Country} & \tiny{RSN} & \tiny{Extended Support Rates} & \tiny{Vendor-Specific} \\ 
   \midrule [0.12em]
   \multirow{4}{*}{\rotatebox{90}{%
  \tiny{Hardware}
}~} 
    \rule{0pt}{1cm} 
   & \tiny{D-Link DIR-615 Wireless-N300 Router}& \tiny{No}& \tiny{\color{red}Yes}& \tiny{\color{red}Yes}& \tiny{Yes}& \tiny{No}& \tiny{No} & \tiny{Yes} & \tiny{Yes} & \tiny{NA} & \tiny{NA} & \tiny{No} & \tiny{No} \\ \cline{2-14} 
    \rule{0pt}{1cm} 
   & \tiny{Digisol DG-HR1400 Wireless Broadband Router}& \tiny{No}& \tiny{\color{red}Yes}& \tiny{\color{red}Yes}& \tiny{\color{red}Yes}& \tiny{No}& \tiny{No} & \tiny{Yes} & \tiny{Yes} & \tiny{NA} & \tiny{NA} & \tiny{No} & \tiny{No}\\
    \cline{2-14}
  \rule{0pt}{1cm}    
& \tiny{TP-Link TL-WR841N 300Mbps Wireless-N Router}& \tiny{No}& \tiny{\color{red}Yes}& \tiny{\color{red}Yes}& \tiny{Yes}& \tiny{Yes} & \tiny{No} & \tiny{Yes} & \tiny{Yes} & \tiny{NA} & \tiny{NA} & \tiny{No} & \tiny{No} \\  \cline{2-14}  
 \rule{0pt}{0.5cm}  
& \tiny{Mi 3C Router}& \tiny{No}& \tiny{\color{red}Yes}& \tiny{\color{red}Yes}& \tiny{Yes}& \tiny{No}& \tiny{No} & \tiny{Yes} & \tiny{No} & \tiny{\color{red}No} & \tiny{NA} & \tiny{No} & \tiny{No}\\ \hline

\multirow{5}{*}{\rotatebox{90}{%
  \tiny{Software}
}~} 
\rule{0pt}{0.5cm}  
& \tiny{hostapd}& \tiny{No}& \tiny{\color{red}Yes}& \tiny{\color{red}Yes}& \tiny{\color{red}Yes}& \tiny{No}& \tiny{\color{red}Yes} & \tiny{Yes} & \tiny{\color{red}Yes} & \tiny{NA} & \tiny{NA} & \tiny{\color{red}No} & \tiny{\color{red}No}\\
 \cline{2-14}
\rule{0pt}{1cm}  
& \tiny{unity network manager} & \tiny{No}& \tiny{\color{red}Yes}& \tiny{\color{red}Yes}& \tiny{Yes}& \tiny{No}& \tiny{No} & \tiny{Yes} & \tiny{No} & \tiny{NA} & \tiny{NA} & \tiny{No} & \tiny{No}\\ \cline{2-14}
 \rule{0pt}{0.5cm}  
 & \tiny{ap-hotspot}& \tiny{No}& \tiny{\color{red}Yes}& \tiny{\color{red}Yes}& \tiny{No}& \tiny{No}& \tiny{No} & \tiny{Yes} & \tiny{No} & \tiny{NA} & \tiny{NA} & \tiny{No} & \tiny{No}\\ \cline{2-14}
 \rule{0pt}{0.5cm}  
  & \tiny{aircrack-ng}& \tiny{No}& \tiny{\color{red}Yes}& \tiny{\color{red}Yes}& \tiny{Yes}& \tiny{No}& \tiny{No} & \tiny{\color{red}No} & \tiny{\color{red}No} & \tiny{\color{red}No} & \tiny{NA} & \tiny{\color{red}No} & \tiny{\color{red}No}\\ \hline
 
 \multirow{5}{*}{\rotatebox{90}{%
  \tiny{Mobile Hotspot}
}~} 
\rule{0pt}{0.5cm}  
   & \tiny{Sony Xperia Z}& \tiny{No}& \tiny{\color{red}Yes}& \tiny{\color{red}Yes}& \tiny{Yes}& \tiny{No} & \tiny{No} & \tiny{No} & \tiny{No} & \tiny{NA} & \tiny{NA} & \tiny{No} & \tiny{No} \\  \cline{2-14}
 \rule{0pt}{0.5cm}  
  & \tiny{Redmi Note 4}& \tiny{No}& \tiny{\color{red}Yes}& \tiny{\color{red}Yes}& \tiny{Yes}& \tiny{Yes} & \tiny{No} & \tiny{No} & \tiny{No} & \tiny{NA} & \tiny{NA} & \tiny{No} & \tiny{No} \\
 \cline{2-14}
 \rule{0pt}{0.5cm}  
 & \tiny{Moto G5 Plus}& \tiny{No}& \tiny{\color{red}Yes}& \tiny{\color{red}Yes}& \tiny{Yes}& \tiny{No}& \tiny{No} & \tiny{No} & \tiny{No} & \tiny{NA} & \tiny{NA} & \tiny{No} & \tiny{No}\\ \cline{2-14}
 \rule{0pt}{0.5cm}  
 & \tiny{Lenovo Tab A7}& \tiny{No}& \tiny{\color{red}Yes}& \tiny{\color{red}Yes}& \tiny{Yes}& \tiny{No}& \tiny{No} & \tiny{No} & \tiny{No} & \tiny{NA} & \tiny{NA} & \tiny{No} & \tiny{No}\\
  
\bottomrule [0.12em]
\end{tabular}
\caption{Experimental results on different devices (hardware/software/mobile hotspots)}
\label{comparisontabletechniques}
\begin{itemize}
\item Yes: The field is \textbf{similar} in beacon frame of ET and AP \texttt{CSE}.
\item No: The field is \textbf{not similar} in beacon frame of ET and AP \texttt{CSE}.
\item {\color{red}Yes}: The field is \textbf{modified} in ET to become similar to AP \texttt{CSE}.
\item {\color{red}No}: The field is \textbf{not present} in legitimate AP \texttt{CSE} but present in ET.
\item NA:- The field in neither present in AP \texttt{CSE} nor in ET of \texttt{CSE}.
\end{itemize}
\end{sidewaystable*}
\indent ETGuard is also tested with the ETs administered in the network by means of mobile hotspots. We establish ETs using Android devices of Sony, Motorola, Lenovo and Redmi. We forged the SSID and BSSID of these hotspots. ETGuard successfully identified the ETs in the network, as the the fields in beacon frames of hotspots and the legitimate one differ (as depicted in Table \ref{comparisontabletechniques}). Mobile hotspot provides the minimum control to users and therefore, modifications of beacon frame fields is not possible.\\
\indent Further, we conducted the experiments for the three attack scenarios (Remote Location/Colocation/Substitution) explained in Subsection \ref{launching}. We analysed that ETGuard effectively handles Colocation attack scenario. For Substitution attack scenario, if an AP is substituted with an ET of different OEM, ETGuard effortlessly detects it. If an AP is replaced with an ET of exactly same OEM and model, but transmits higher signal strength, ETGuard is capable of detecting it. However, if the AP is replaced with an ET of exactly same OEM and model, and transmits precisely similar signal strength, ETGuard may not be able to detect it. In the case of Remote Location attack scenario, genuine AP is not present in the network. The attacker can locate ET at any location in the network. ETGuard can detect the ET in every situation, unless an attacker situates an ET of exactly same OEM and model, and even manages to transmit similar signal strength as of original AP. This is an extremely rare scenario because determining the OEM, model and strength for an AP that has been removed from the network is an extremely challenging task. So, we rule out this possibility. Notably, ETGuard maintains the database entries for removed APs to prevent Remote Location attack.     
\subsubsection{Accuracy}
To assess the accuracy of ETGuard, we discuss the false-positives and false-negatives incurred by ETGuard. In this work, the positive and negative factors are defined as:
\begin{itemize}
\item \textit{True-Positive (TP)}: If the AP is genuine, and ETGuard identifies it as a genuine AP, the result is considered as TP.
\item \textit{False-Positive (FP)}: If the AP is an ET introduced in the network, and ETGuard identifies it as a genuine AP, the result is considered as FP.
\item \textit{True-Negative (TN)}: If the AP is an ET introduced in the network, and ETGuard detects it as an ET, the result is considered as TN.
\item \textit{False-Negative (FN)}: If the AP is genuine, and ETGuard detects it as an ET, the result is considered as FN.  
\end{itemize}  
ETGuard incurs only one case of \textit{FP}, when an attacker launches a Substitution attack scenario with an ET of exactly same OEM and model, and transmits identical signal strength as of genuine AP. When an ET of similar OEM and model (as that of original AP) is launched in the network, the beacon frame components of the ET are analogous to the genuine AP. So, the first attempt of ETGuard to match the fingerprints is unable to identify the ET (see the Algorithm \ref{hamming}). Further, ETGuard compares the stored signal strength of original AP with the ET. If the attacker maintains the signal strength similar to the original one, ETGuard incurs \textit{FP}. However, ETGuard incurs no \textit{FNs}. The reason for obtaining only one case of \textit{FP} and no cases of \textit{FN} is the difference in various hardware and driver dependent fields in beacon frames such as TIM, Capability Information, Supported Rates, Vendor-Specific, and RSN settings (The reasons for difference in such fields are explained in Subsection \ref{serverana}). Further, signal strength element of beacon frames is a network dependent component. Since these fields are not modifiable, and also vary from OEM to OEM for hardware AP, from hardware AP to software AP, from hardware AP to mobile hotspot, and from mobile hotspot to software AP, thus, ETGuard is capable of successfully detecting all types of ETs and ET attack scenarios with very less \textit{FPs} and no \textit{FNs}.   
\subsection{Client App Interface for ETGuard}
\label{client}
We provide an app interface for users which on contrary to Android WI-FI settings, displays SSID, BSSID and signal strength of the AP. This app initially sends request to the server for network scanning, and on receiving the response, it refreshes the WI-FI list and highlights the ETs with red color to assist user in connecting with the legitimate AP.
\subsection{Discussion}
ETGuard experiences a delay between receiving the client request and responding with the ETs present in the network. ETGuard comprises of $7$ modules (as explained in Section \ref{approach}). The delay is incurred at the Packet Handler module, as it captures beacon frames on $13$ channels in the network (as explained in Subsection \ref{implementation}). The duration to capture the frames on the channels can be adjusted based on the location of ETGuard in the network. If the APs are in a clear range of ETGuard, duration can be reduced, or else frames need to be captured for a longer duration. Since an AP transmit a beacon frame every $102.4ms$\cite{mgmt}, the duration for capturing beacon frames should be greater than $102.4ms$. Further, the Extractor module extracts the relevant fields from the beacon frames one by one, and forwards it to the Processing Module. Since the fields extracted for fingerprinting are not arranged sequentially in the beacon frames, and also the size of fields is not constant for each beacon frame, this module incurs a delay. In our experiments, Extractor module of ETGuard incurred a delay of $650ms$. Further, the Processing Module and Fingerprint Storage Engine module incurs a very negligible delay, as mysql processing is extremely fast. Thus, ETGuard incurs a delay of few seconds.\\
\indent ETGuard is easy to deploy in any network, as only $1$ out of $7$ modules of ETGuard needs modification, i.e., Fingerprint Storage Engine requires modification. Since Fingerprint Storage Engine stores the fingerprints of the APs present in the network, this module varies from network to network, and needs to be updated for every network. Rest all the modules are not network dependent. Further, ETGuard is proposed to detect the presence of ETs for infrastructure based networks, i.e., local networks. To extend the deployment of ETGuard for global networks, multiple instances of ETGuard needs to be installed, depending on the physical range of the network. Let us assume that the global network contains multiple APs of similar OEM, model, and configured with similar SSID. The fingerprints of these APs differ, as some of the beacon frame components among such APs are different, such as MAC address and signal strength. Thus, only the Fingerprint Storage Engine module needs modification. Thus, ETGuard is practically feasible and easy to deploy in any network setting. Notably, other than MAC address and signal strength fields of beacon frame, all the other fields used for fingerprinting are network and region independent, and therefore, does not affect the fingerprints, and hence, the deployment. 
\subsection{Comparison with Existing Techniques}
\begin{sidewaystable*}[htbp]
  \centering
  \begin{tikzpicture}
 \filldraw[color=black, fill=white, very thick](0,0) circle (0.13); \end{tikzpicture} = The technique ideally handles it, \begin{tikzpicture}
 \filldraw[color=black, fill=black, very thick](0,0) circle (0.13); \end{tikzpicture} = The technique uses it but it is not ideal for a practical solution, \begin{tikzpicture}
 \filldraw[color=black, fill=black, very thick](0,0) rectangle (0.2,0.2); \end{tikzpicture} = The technique does not handle it, but an ideal solution should handle it, \begin{tikzpicture}
 \filldraw[color=black, fill=white, very thick](0,0) rectangle (0.2,0.2); \end{tikzpicture} = The technique does not ideally uses it and \statcirc[black]{white} = The technique cannot handle all the cases of it. 
 \footnotesize
  \begin{tabular}{LNNNNNNNNNN}
   \hline
   \hline
   \textbf{S. No.} & \textbf{Technique} & \textbf{Pre-Association} & \textbf{H/W} & \textbf{S/W} & \textbf{Hotspot} & \textbf{Remote Location} & \textbf{Colocation} & \textbf{Substitution} & \textbf{Extra H/w Utilised} & \textbf{Protocol Modification} \\
    \hline
    \hline
    \rule{0pt}{0.4cm} 
    1 & {\color{red}ETGuard} & \begin{tikzpicture}
\filldraw[color=black, fill=white, very thick](0,0) circle (0.13); \end{tikzpicture} & \begin{tikzpicture}
\filldraw[color=black, fill=white, very thick](0,0) circle (0.13); \end{tikzpicture} & \begin{tikzpicture}
\filldraw[color=black, fill=white, very thick](0,0) circle (0.13); \end{tikzpicture} & \begin{tikzpicture}
\filldraw[color=black, fill=white, very thick](0,0) circle (0.13); \end{tikzpicture} & \begin{tikzpicture}
\filldraw[color=black, fill=white, very thick](0,0) circle (0.13); \end{tikzpicture} & \begin{tikzpicture}
\filldraw[color=black, fill=white, very thick](0,0) circle (0.13); \end{tikzpicture} & \statcirc[black]{white} & \begin{tikzpicture}
\filldraw[color=black, fill=white, very thick](0,0) rectangle (0.2,0.2); \end{tikzpicture} & \begin{tikzpicture}
\filldraw[color=black, fill=white, very thick](0,0) rectangle (0.2,0.2); \end{tikzpicture} \\
\hline
\rule{0pt}{0.4cm} 
    2 & DRET\cite{tang2017exploiting} & \begin{tikzpicture}
\filldraw[color=black, fill=white, very thick](0,0) circle (0.13); \end{tikzpicture} & \begin{tikzpicture}
\filldraw[color=black, fill=white, very thick](0,0) circle (0.13); \end{tikzpicture} & \begin{tikzpicture}
\filldraw[color=black, fill=black, very thick](0,0) rectangle (0.2,0.2); \end{tikzpicture} & \begin{tikzpicture}
\filldraw[color=black, fill=black, very thick](0,0) rectangle (0.2,0.2); \end{tikzpicture} & \begin{tikzpicture}
\filldraw[color=black, fill=white, very thick](0,0) circle (0.13); \end{tikzpicture} & \begin{tikzpicture}
\filldraw[color=black, fill=black, very thick](0,0) rectangle (0.2,0.2); \end{tikzpicture} & \begin{tikzpicture}
\filldraw[color=black, fill=black, very thick](0,0) rectangle (0.2,0.2); \end{tikzpicture} & \begin{tikzpicture}
\filldraw[color=black, fill=white, very thick](0,0) rectangle (0.2,0.2); \end{tikzpicture} & \begin{tikzpicture}
\filldraw[color=black, fill=white, very thick](0,0) rectangle (0.2,0.2); \end{tikzpicture} \\
\hline
\rule{0pt}{0.4cm} 
	3 & ET Detector\cite{hsu2017client} & \begin{tikzpicture}
\filldraw[color=black, fill=black, very thick](0,0) rectangle (0.2,0.2); \end{tikzpicture} & \statcirc[black]{white} & \begin{tikzpicture}
\filldraw[color=black, fill=white, very thick](0,0) circle (0.13); \end{tikzpicture} & \begin{tikzpicture}
\filldraw[color=black, fill=black, very thick](0,0) rectangle (0.2,0.2); \end{tikzpicture} & \begin{tikzpicture}
\filldraw[color=black, fill=black, very thick](0,0) circle (0.13); \end{tikzpicture} & \begin{tikzpicture}
\filldraw[color=black, fill=white, very thick](0,0) circle (0.13); \end{tikzpicture} & \begin{tikzpicture}
\filldraw[color=black, fill=black, very thick](0,0) rectangle (0.2,0.2); \end{tikzpicture} & \begin{tikzpicture}
\filldraw[color=black, fill=white, very thick](0,0) rectangle (0.2,0.2); \end{tikzpicture} & \begin{tikzpicture}
\filldraw[color=black, fill=white, very thick](0,0) rectangle (0.2,0.2); \end{tikzpicture} \\
\hline
\rule{0pt}{0.6cm} 
	4 & Hacker's Toolbox\cite{lanze2015hacker} & \begin{tikzpicture}
\filldraw[color=black, fill=white, very thick](0,0) circle (0.13); \end{tikzpicture} & \begin{tikzpicture}
\filldraw[color=black, fill=black, very thick](0,0) rectangle (0.2,0.2); \end{tikzpicture} & \begin{tikzpicture}
\filldraw[color=black, fill=white, very thick](0,0) circle (0.13); \end{tikzpicture} & \begin{tikzpicture}
\filldraw[color=black, fill=black, very thick](0,0) rectangle (0.2,0.2); \end{tikzpicture} & \begin{tikzpicture}
\filldraw[color=black, fill=white, very thick](0,0) circle (0.13); \end{tikzpicture} & \begin{tikzpicture}
\filldraw[color=black, fill=white, very thick](0,0) circle (0.13); \end{tikzpicture} & \begin{tikzpicture}
\filldraw[color=black, fill=white, very thick](0,0) circle (0.13); \end{tikzpicture} & \begin{tikzpicture}
\filldraw[color=black, fill=white, very thick](0,0) rectangle (0.2,0.2); \end{tikzpicture} & \begin{tikzpicture}
\filldraw[color=black, fill=white, very thick](0,0) rectangle (0.2,0.2); \end{tikzpicture} \\
\hline
\rule{0pt}{0.4cm} 
	5 & CETAD\cite{mustafa2014cetad} & \begin{tikzpicture}
\filldraw[color=black, fill=black, very thick](0,0) rectangle (0.2,0.2); \end{tikzpicture} & \begin{tikzpicture}
\filldraw[color=black, fill=white, very thick](0,0) circle (0.13); \end{tikzpicture} & \begin{tikzpicture}
\filldraw[color=black, fill=white, very thick](0,0) circle (0.13); \end{tikzpicture} & \begin{tikzpicture}
\filldraw[color=black, fill=white, very thick](0,0) circle (0.13); \end{tikzpicture} & \begin{tikzpicture}
\filldraw[color=black, fill=white, very thick](0,0) circle (0.13); \end{tikzpicture} & \begin{tikzpicture}
\filldraw[color=black, fill=white, very thick](0,0) circle (0.13); \end{tikzpicture} & \begin{tikzpicture}
\filldraw[color=black, fill=black, very thick](0,0) rectangle (0.2,0.2); \end{tikzpicture} & \begin{tikzpicture}
\filldraw[color=black, fill=white, very thick](0,0) rectangle (0.2,0.2); \end{tikzpicture} & \begin{tikzpicture}
\filldraw[color=black, fill=white, very thick](0,0) rectangle (0.2,0.2); \end{tikzpicture} \\
\hline
\rule{0pt}{0.4cm} 
    6 & TSF\cite{lanze2014undesired} & \begin{tikzpicture}
\filldraw[color=black, fill=white, very thick](0,0) circle (0.13); \end{tikzpicture} & \begin{tikzpicture}
\filldraw[color=black, fill=black, very thick](0,0) rectangle (0.2,0.2); \end{tikzpicture} & \begin{tikzpicture}
\filldraw[color=black, fill=white, very thick](0,0) circle (0.13); \end{tikzpicture} & \begin{tikzpicture}
\filldraw[color=black, fill=black, very thick](0,0) rectangle (0.2,0.2); \end{tikzpicture} & \begin{tikzpicture}
\filldraw[color=black, fill=white, very thick](0,0) circle (0.13); \end{tikzpicture} & \begin{tikzpicture}
\filldraw[color=black, fill=white, very thick](0,0) circle (0.13); \end{tikzpicture} & \begin{tikzpicture}
\filldraw[color=black, fill=white, very thick](0,0) circle (0.13); \end{tikzpicture} & \begin{tikzpicture}
\filldraw[color=black, fill=black, very thick](0,0) circle (0.13); \end{tikzpicture} & \begin{tikzpicture}
\filldraw[color=black, fill=white, very thick](0,0) rectangle (0.2,0.2); \end{tikzpicture} \\
\hline
\rule{0pt}{0.6cm} 
	7 & ClockSkew + Temperature\cite{lanze2014letting} & \begin{tikzpicture}
\filldraw[color=black, fill=white, very thick](0,0) circle (0.13); \end{tikzpicture} & \begin{tikzpicture}
\filldraw[color=black, fill=white, very thick](0,0) circle (0.13); \end{tikzpicture} & \begin{tikzpicture}
\filldraw[color=black, fill=black, very thick](0,0) rectangle (0.2,0.2); \end{tikzpicture} & \begin{tikzpicture}
\filldraw[color=black, fill=black, very thick](0,0) rectangle (0.2,0.2); \end{tikzpicture} & \begin{tikzpicture}
\filldraw[color=black, fill=white, very thick](0,0) circle (0.13); \end{tikzpicture} & \begin{tikzpicture}
\filldraw[color=black, fill=white, very thick](0,0) circle (0.13); \end{tikzpicture} & \begin{tikzpicture}
\filldraw[color=black, fill=white, very thick](0,0) circle (0.13); \end{tikzpicture} & \begin{tikzpicture}
\filldraw[color=black, fill=black, very thick](0,0) circle (0.13); \end{tikzpicture} & \begin{tikzpicture}
\filldraw[color=black, fill=white, very thick](0,0) rectangle (0.2,0.2); \end{tikzpicture} \\
\hline
\rule{0pt}{0.6cm} 
	8 & Inter-packet arrival time\cite{shetty2007rogue} & \begin{tikzpicture}
\filldraw[color=black, fill=white, very thick](0,0) circle (0.13); \end{tikzpicture} & \begin{tikzpicture}
\filldraw[color=black, fill=white, very thick](0,0) circle (0.13); \end{tikzpicture} & \begin{tikzpicture}
\filldraw[color=black, fill=black, very thick](0,0) rectangle (0.2,0.2); \end{tikzpicture} & \begin{tikzpicture}
\filldraw[color=black, fill=black, very thick](0,0) rectangle (0.2,0.2); \end{tikzpicture} & \begin{tikzpicture}
\filldraw[color=black, fill=white, very thick](0,0) circle (0.13); \end{tikzpicture} & \begin{tikzpicture}
\filldraw[color=black, fill=white, very thick](0,0) circle (0.13); \end{tikzpicture} & \begin{tikzpicture}
\filldraw[color=black, fill=white, very thick](0,0) circle (0.13); \end{tikzpicture} & \begin{tikzpicture}
\filldraw[color=black, fill=white, very thick](0,0) rectangle (0.2,0.2); \end{tikzpicture} & \begin{tikzpicture}
\filldraw[color=black, fill=white, very thick](0,0) rectangle (0.2,0.2); \end{tikzpicture} \\
\hline
\rule{0pt}{0.4cm} 
	9 & WifiHop\cite{monica2011wifihop} & \begin{tikzpicture}
\filldraw[color=black, fill=black, very thick](0,0) rectangle (0.2,0.2); \end{tikzpicture} & \begin{tikzpicture}
\filldraw[color=black, fill=white, very thick](0,0) circle (0.13); \end{tikzpicture} & \begin{tikzpicture}
\filldraw[color=black, fill=black, very thick](0,0) rectangle (0.2,0.2); \end{tikzpicture} & \begin{tikzpicture}
\filldraw[color=black, fill=black, very thick](0,0) rectangle (0.2,0.2); \end{tikzpicture} & \begin{tikzpicture}
\filldraw[color=black, fill=black, very thick](0,0) rectangle (0.2,0.2); \end{tikzpicture} & \begin{tikzpicture}
\filldraw[color=black, fill=white, very thick](0,0) circle (0.13); \end{tikzpicture} & \begin{tikzpicture}
\filldraw[color=black, fill=black, very thick](0,0) rectangle (0.2,0.2); \end{tikzpicture} & \begin{tikzpicture}
\filldraw[color=black, fill=white, very thick](0,0) rectangle (0.2,0.2); \end{tikzpicture} & \begin{tikzpicture}
\filldraw[color=black, fill=white, very thick](0,0) rectangle (0.2,0.2); \end{tikzpicture} \\
\hline
\rule{0pt}{0.4cm} 
	10 & SOWA\cite{byrd2011secure} & \begin{tikzpicture}
\filldraw[color=black, fill=white, very thick](0,0) circle (0.13); \end{tikzpicture} & \begin{tikzpicture}
\filldraw[color=black, fill=white, very thick](0,0) circle (0.13); \end{tikzpicture} & \begin{tikzpicture}
\filldraw[color=black, fill=black, very thick](0,0) rectangle (0.2,0.2); \end{tikzpicture} & \begin{tikzpicture}
\filldraw[color=black, fill=black, very thick](0,0) rectangle (0.2,0.2); \end{tikzpicture} & \begin{tikzpicture}
\filldraw[color=black, fill=white, very thick](0,0) circle (0.13); \end{tikzpicture} & \begin{tikzpicture}
\filldraw[color=black, fill=white, very thick](0,0) circle (0.13); \end{tikzpicture} & \begin{tikzpicture}
\filldraw[color=black, fill=white, very thick](0,0) circle (0.13); \end{tikzpicture} & \begin{tikzpicture}
\filldraw[color=black, fill=white, very thick](0,0) rectangle (0.2,0.2); \end{tikzpicture} & \begin{tikzpicture}
\filldraw[color=black, fill=black, very thick](0,0) circle (0.13); \end{tikzpicture} \\
\hline
\rule{0pt}{0.4cm} 
	11 & ETSniffer\cite{yang2012active} & \begin{tikzpicture}
\filldraw[color=black, fill=black, very thick](0,0) rectangle (0.2,0.2); \end{tikzpicture} & \statcirc[black]{white} & \begin{tikzpicture}
\filldraw[color=black, fill=white, very thick](0,0) circle (0.13); \end{tikzpicture} & \begin{tikzpicture}
\filldraw[color=black, fill=black, very thick](0,0) rectangle (0.2,0.2); \end{tikzpicture} & \begin{tikzpicture}
\filldraw[color=black, fill=black, very thick](0,0) rectangle (0.2,0.2); \end{tikzpicture} & \begin{tikzpicture}
\filldraw[color=black, fill=white, very thick](0,0) circle (0.13); \end{tikzpicture} & \begin{tikzpicture}
\filldraw[color=black, fill=black, very thick](0,0) rectangle (0.2,0.2); \end{tikzpicture} & \begin{tikzpicture}
\filldraw[color=black, fill=white, very thick](0,0) rectangle (0.2,0.2); \end{tikzpicture} & \begin{tikzpicture}
\filldraw[color=black, fill=white, very thick](0,0) rectangle (0.2,0.2); \end{tikzpicture} \\
\hline
\hline
  \end{tabular}
  \caption{Comparison of ETGuard with the existing state-of-the-art approaches} 
  \label{compare}
\end{sidewaystable*}
We have compared ETGuard with existing state-of-the-art approaches on the basis of various parameters, such as whether the approach could detect an ET before association or not, whether the approach can detect all the three types of ETs (hardware/software/mobile hotspot) or not, whether the approach handles all the three types of attack scenarios (Remote Location/Colocation/Substitution) or not, whether the approach utilises any extra hardware for analysis or not, and whether the approach modifies any protocols for ET detection or not. The comparison is illustrated in Table \ref{compare}. The less the number of filled circle or square (\begin{tikzpicture}
\filldraw[color=black, fill=black, very thick](0,0) circle (0.13); \end{tikzpicture} and \begin{tikzpicture}
\filldraw[color=black, fill=black, very thick](0,0) rectangle (0.2,0.2); \end{tikzpicture} respectively) in the row of a particular technique, the more effective the technique is for ET detection. According to Table \ref{compare}, DRET\cite{tang2017exploiting} cannot detect ETs created through softwares and mobile hotspots, and unable to handle Colocation and Substitution attack scenarios for ET launch, because it uses SSI to locate only stationary ETs, and they assume that attacker launches an ET at a remote location from the genuine AP. Hacker's Toolbox\cite{lanze2015hacker} can only detect software ETs. Similarly, TSF\cite{lanze2014undesired} solely handles software APs. However it utilises an extra monitoring device which makes it infeasible for the real environment. Likewise, ClockSkew + Temperature\cite{lanze2014letting} also utilises an extra hardware for ET detection, and can only deal with hardware ETs. Furthermore, Inter-packet arrival time\cite{shetty2007rogue} cannot identify ETs launched through softwares and mobile hotspots.  CETAD\cite{mustafa2014cetad}, ETSniffer\cite{yang2012active}, ET Detector\cite{hsu2017client} and WifiHop\cite{monica2011wifihop} are not pre-association techniques.  SOWA\cite{byrd2011secure} incorporates protocol modification to detect hardware ETs. They have incorporated digital certificate based unique identification for APs. However, it requires a central certification authority and complete protocol modification, and thus, cannot be deployed in real environment. Our proposed approach named as ETGuard is a pre-association technique, can handle all types of ETs, does not modify any protocols, does not enforce any additional expensive hardware, and handles all attack scenarios with a limitation in Substitution attack scenario. ETGuard can not handle a case where an attacker has substituted the AP with an ET of same OEM and model, and transmits signal strength similar to that of AP. However, this is a rarest scenario. Thus, we can assert from the comparison in Table \ref{compare} that ETGuard outperforms the existing approaches for ET detection.   
\section{Related Work}
\label{related}
In this section, we discuss the existing literature of two realms - D2D attacks on Android and ET detection techniques.
\subsection{D2D Attacks on Android}
Since Android is based on Linux kernel, it inherits Linux kernel features, either without modification, or with slight alteration. As Linux is vulnerable to certain D2D attacks such as MITM, sniffing, replay attacks, etc., Android too becomes susceptible to these attacks. The most recent attack on Android is the Key Reinstallation Attack (KRACK)\cite{vanhoef2017key}. This attack exploits the 4-way handshake mechanism of wpa\_supplicant to bluff Android devices to reuse nonces, and reinstall an all zero encryption key. Thus, the attacker can either replay, or decrypt, or forge the data frames. Another popular attack on Android devices includes MITM attack. 
MalloDroid\cite{fahl2012eve} statically analyses Android apps for improper and vulnerable usage of SSL by inspecting the code for SSL usage, hostname verification and certificate trust APIs. 
However, MalloDroid does not conduct data flow analysis, and hence, may incur a lot of false positives. Another exploitable feature acquired by Android from Linux includes socket implementation for device to device communication over the network. 
OPAnalyzer\cite{jia2017open} statically analyses Android apps for open port vulnerabilities. It takes into account three classes of adversaries - malicious app on the same device, local network attacker and malicious scripts on the web. It conducts entry point analysis, followed by taint analysis and reachability analysis to confirm the vulnerabilities in an app. However, it only handles TCP sockets, and does not consider sensitive information transmission from native layer to java layer.   
\subsection{ET Detection}
We classified the existing solutions in two categories - pre-association and post-association. The post-association techniques require connectivity with the AP for the detection of ET, and on the contrary, pre-association techniques are capable  of detecting ETs before the connection and transmission of data traffic. 
\subsubsection{Pre-association}
SOWA\cite{byrd2011secure} implements protocol modification for ensuring protection against ETs. It ties the SSID with a certificate issued by a valid certification authority, and checks the certificate of an AP before connecting to it. However, the solution is not feasible as the deployment needs the firmware and drivers of APs and clients to get updated. Neumann et al.\cite{neumann2012empirical} utilises the inter-packet arrival time to identify ETs. They construct histogram for the frame arrival times of individual APs. However, during implementation, the detection rate falls to 50-60\%, and thus, not a feasible solution. Some approaches construct hardware fingerprints using obligatory physical phenomenon such as clock skew. Jana et al.\cite{jana2010fast} utilises the clock skew phenomenon to create hardware fingerprints for an AP. The clock skew induces clocks based on crystal oscillators to possess slight alterations in speed. However, many APs reflected similar alterations, and thus, the technique fails in efficiently detecting ETs. Further, Lanze et al.\cite{lanze2014letting} improved the clock skew based technique by including underlying device dependency on temperature to detect ETs. DRET\cite{tang2017exploiting} utilises radiometric signal properties to locate ETs. The technique is based on the hypothesis that the RSSI values of an AP located at a particular place is always same. However, it does not consider the fact that RSSI values oscillate due to hazy effects, and are not always the same. Unfortunately, the above techniques can only detect ETs created through hardwares. Lanze et al.\cite{lanze2014undesired} proposed a dedicated approach for detection of software APs by using Timing Synchronization Functions (TSF). They recorded the timestamps for beacon transmissions of software APs, and plotted them. According to them, hardware APs always form a linear pattern, whereas, software APs possess outliers. None of the pre-association techniques are suitable for detecting ETs constructed through mobile hotspots. 
\subsubsection{Post-association}
ETSniffer\cite{yang2012active} proposed two algorithms - Training Mean Matching (TMM) and Hop Differentiating Technique (HDT) for detection of ETs. TMM employs Inter-packet arrival time as the differentiating characteristic between genuine and fake AP. It considers the fact that genuine AP follows one-hop network topology, whereas, an ET follows multi-hop topology, as an ET uses genuine AP to provide Internet connectivity. TMM trains the values of Inter-packet arrival time for one hop and multi-hop topology. HDT improves TMM by removing the need of training because HDT analyses Server-to-AP Inter-packet arrival time ratio and assumes a theoretical value for the topologies. Similarly, WifiHop\cite{monica2011wifihop} also considers difference in the network topology of an AP and an ET as a detection parameter. They transmit watermarked packets on the network before the user traffic to check whether the packets are transmitted on the correct route or not. Further, ET Detector\cite{hsu2017client} also monitors the behavior of packets in the network to detect ETs. They assume the fact that ET will transmit the packet to the real AP to forward it further. CETAD\cite{mustafa2014cetad} detects an ET on the basis of three parameters - difference in the Internet Service Provider, RTT and the standard deviation of RTT. The post-association techniques are incapable of preventing Invoking Malicious Component attack (explained in Section \ref{threat}) launched through ETs.\\
\section{Conclusion and Future Work}
\label{Conclusion}
In this paper, we demonstrate the impact of D2D attacks on Android devices, launched in a network using ETs. We illustrate the launch and adverse effect of the ``Invoking Malicious Component" attack on Android. The attack possesses the capability of inflicting an Android device before the connection and transmission of data traffic through an ET. The contemporary ET detection solutions are incapable of preventing this attack because either they analyse an ET after the relay of user traffic through it, or they can detect this attack only for hardware ETs. We propose an automated, online, incremental, fingerprinting based pre-association technique known as ETGuard, which utilises beacon frames to fingerprint an AP. The fingerprints are stored on the server which is requested by the client app installed on an Android device to scan the network for ETs in real-time. The advantages of ETGuard are manifold - it does not need any expertise for deployment, no expensive hardware is used and no protocols are adapted. 
To assess the performance of ETGuard, we deploy it in real network, and launch ETs using hardware, software and hotspots. ETGuard successfully identifies all scenarios of ET launch with no false negatives, but incurred false positives in one scenario when an attacker substitutes an AP with an ET of exactly same OEM and model, and transmits signal strength similar to that of original AP. In future, we plan to extend ETGuard for handling the Substitution attack scenario for ET and AP belonging to the same OEM. We also plan to offload the detection mechanism on Android devices, instead of following the client-server architecture.
\balance
\section*{References}
\bibliographystyle{elsarticle-num}
\biboptions{numbers,sort&compress}
\bibliography{eviltwin}
\end{document}